\pdfoutput=1
\documentclass[%
reprint,
superscriptaddress,
 amsmath,amssymb,
 aps,
]{revtex4-1}
\usepackage{graphicx}% Include figure files
\usepackage{dcolumn}% Align table columns on decimal point
\usepackage{bm}% bold math
\usepackage[colorlinks, linkcolor=blue, urlcolor=black, citecolor=blue]{hyperref}
\begin{document}

\preprint{APS/123-QED}

\title{Radiation induced oscillating gap states of nonequilibrium two-dimensional superconductors}% Force line breaks with \\
\
\author{Huiying Liu}
 \affiliation{%
International Center for Quantum Materials, Peking University, Beijing 100871, China}
\author{Junren Shi}
 \affiliation{%
International Center for Quantum Materials, Peking University, Beijing 100871, China}
 \affiliation{%
Collaborative Innovation Center of Quantum Matter, Beijing 100871, China}
\date{\today}% It is always \today, today,

\begin{abstract}
We study effects of infrared radiations on a two-dimensional BCS superconductor coupled
with a normal metal substrate through a tunneling barrier. The phase transition conditions
are analyzed by inspecting stability of the system against perturbations of pairing potentials. 
We find an oscillating gap phase 
with a frequency not directly related to the radiation frequency but
resulting from the asymmetry of electron density of states of the
system as well as the tunneling amplitude. When such a superconductor is in contact with another superconductor,
it will give rise to an unusual alternating Josephson current.  

\begin{description}
\item[PACS numbers]
\verb+74.40.Gh+, \verb+74.25.N-+, \verb+74.78.-w+, \verb+74.50.+r+

\end{description}
\end{abstract}

\pacs{Valid PACS appear here}% PACS, the Physics and Astronomy
\maketitle

%\tableofcontents

%\section{introduction}
The rapid development of time-resolved
spectroscopy technology have drawn growing interests in the study of nonequilibrium phenomena. The conductivity properties of solid materials can be greatly changed in nonequlibrium states induced by radiation.
 In two-dimensional electron gas, radiation induces zero resistance states at  high Landau filling factors
\cite{zudov2003evidence,mani2002zero,shi2003radiation,durst2003radiation}. Recent discoveries on high temperature cuprate superconductors
reveal that infrared radiations transform nonsuperconducting
compounds into transient superconductors or enhance coherent superconducting
transport even at temperature above the superconducting transition
temperature  \cite{fausti2011light,hu2014optically}.
For irradiated BCS superconductors, the enhancement of superconducting gap  \cite{wyatt1966mic, dayem1967behav, Kom1977meas, beck2013transi} or the oscillating amplitude modes in nonadiabatic regime \cite{bara2004collec, bara2006synch, yuz2005noneq, yuz2006relax, sch2011reson, matsu2013higgs} induced at near gap frequencies have been discussed in many previous works.
Recently, experimental studies on two-dimensional superconductors 
have shown superconducting order remains robust in ultra thin crystalline films which are a few atomic layers thick \cite{qin2009superconductivity,zhang2010superconductivity,guo2004superconductivity,ozer2006hard,ozer2007tuning}. 
It brings new opportunities to the study of non-equilibrium superconducting orders
in reduced dimensions.
 
 %The enhancement of superconducting gap in irradiated BCS superconductors at near gap frequencies has also been observed \cite{wyatt1966mic,*dayem1967behav,*Kom1977meas,*beck2013transi}. %Radiation-induced effects also bring new opportunities to the study of superconductivity. 

In this Letter, we investigate
 effects of radiations on a two-dimensional BCS superconductor
coupled with a normal metal through a tunneling barrier and find
a new oscillating state induced by radiations which has not been discussed
before. Different from the usual (fractional) ac Josephson effect \cite{josephson1962possible,shapiro1963josephson}, in which oscillation is induced by dc voltages, the frequency of this alternating phase is not directly affected
by the radiation frequency $f_{0}$, but the radiation intensity,
the amplitude of the tunneling interaction and most importantly, the
 asymmetry of density of states around Fermi-level. Such a state can
be a probe of the internal properties of 2D superconductors.

%\section{Theory}

\begin{figure}[tb!]
\includegraphics[width=6cm]{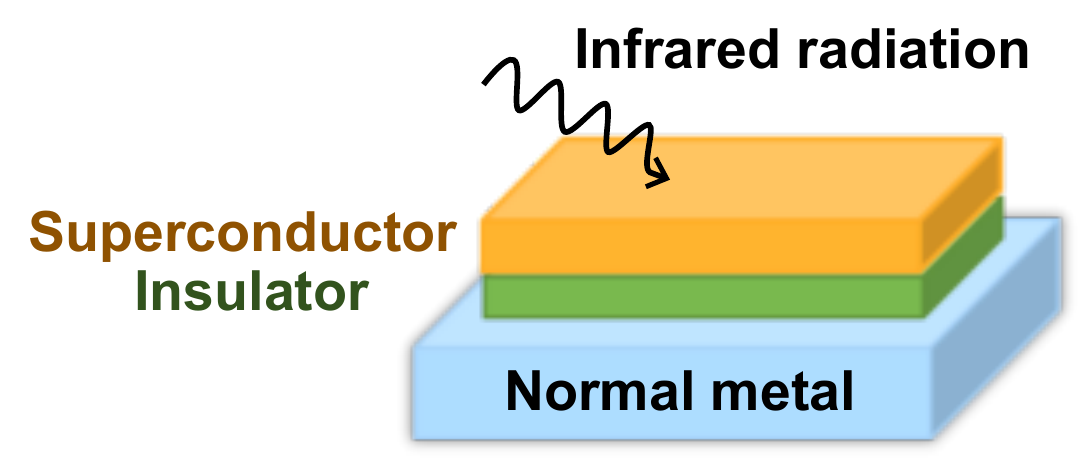}% Here is how to import EPS art
\caption{\label{fig_1} A schematic diagram of the system. The superconducting film, coupled with the metal substrate by a insulating film, is irradiated by infrared wave whose wavelength is resonant with a phonon mode in the superconductor. }
\end{figure}

The physical system we concern is a superconducting film coupled to
a normal metal substrate by a tunneling barrier. The amplitude of the tunneling matrix element is determined
by the thickness of the insulating film. We excite the superconductor
with infrared radiations, the wavelength of which is assumed to be resonant with
a phonon mode in the superconductor (shown in Fig.\ref{fig_1}).
As a result, the electrons feel an oscillating
crystal field resulting from the excited phonons. The Hamiltonian
of this tunneling system can be written as the sum of the Hamiltonians
of N, S and the tunneling barrier, 
\setlength\abovedisplayskip{5pt} 
\setlength\belowdisplayskip{5pt} 
{\setlength{\arraycolsep}{-1pt}
\begin{gather} 
H=H_{N}+H_{S}+H_{T} \notag \\
H_{S}\!= \!\underset{k}{\sum}\Psi_{k}^{\dagger}\!\left(\begin{array}{cc}
\tilde{\varepsilon}_{k}+V_{k}\cos(\omega_{0}t) & 0\\
0 & -\tilde{\varepsilon}_{k}-V_{k}\cos(\omega_{0}t)
\end{array}\right)\!\Psi_{k} \!+\!V_{int}\notag \\
H_{T}=\sum_{k,k'}T_{k,k'}\left(c_{k\uparrow}^{\dagger}d_{k'\uparrow}+c_{-k\uparrow}d_{-k'\uparrow}^{\dagger}\right)+h.c.\label{eq_Ham}
\end{gather}}where
 $c_{k}$ and $d_{k'}$ are annihilation operators of the single
particle state in the superconductor and the normal metal substrate
respectively. The Hamiltonian of superconductor $H_{S}$ with a single
frequency oscillating field $V_{k}\cos(\omega_{0}t)$ is written in the Nambu matrix form \cite{nambu1961dynamical,*gor1959microscopic}, where
$\tilde{\varepsilon}=\varepsilon-\mu$ and $\Psi_{k}=\bigl[\begin{array}{cc}
c_{k\uparrow} & c_{-k\downarrow}^{\dagger}\end{array}\bigr]^\mathrm{ T }$ is the basis of Nambu representation, and 
$V_{int}$ is the electron-electron interaction. $T_{k,k'}$ is the tunneling amplitude between S and N. For simplicity, $\hbar$ and $e$ are omitted in the discussion.

To determine superconducting phase transition of the irradiated system,
we apply the theory of linear response and investigate the system's
stability against a weak external paring potential $\Delta_{ext}$$\left(t\right)$
which perturbs the system with $V_{ext}=\Delta_{ext}c_{k\uparrow}^{\dagger}c_{-k\downarrow}^{\dagger}+\Delta_{ext}^{*}c_{-k\downarrow}c_{k\uparrow}$.
The response function $\chi_{k,k'}\left(t,t'\right)$ of paring amplitude
$\psi_{k}=\left\langle c_{-k\downarrow}\left(t\right)c_{k\uparrow}\left(t\right)\right\rangle $
to the perturbation is defined by 
\begin{equation}\label{chi}
\psi_{k}\left(t\right)=\sum_{k'}\int \mathrm{d}t'\chi_{k,k'}\left(t,t'\right)\Delta_{ext}\left(k',t'\right)
\end{equation}
In the presence of external paring potential, the disturbance will reach
a steady state in the normal state, while in the superconducting state,
however small perturbation can lead to a disturbance
increasing with the time. Superconductivity is a result of the instability
of the system to the perturbing pairing potential.

To determine the response function of the system, we employ the RPA-like
approach. First, we define $\chi_{0\,k,k'}\left(t,t'\right)_{}$ as the
pairing response function of a noninteracting system, i.e. $V_{int}=0$
in Eq.~(\ref{eq_Ham}). When the interaction is included, an electron in the system
will feel not only the external pairing potential, but also an induced
pairing potential $\Delta_{ind}\left(t\right)$ exerted by all other
electrons in the system. As a result: 
\begin{equation}\label{chi0}
\psi_{k}\left(t\right)\!=\!\sum_{k'}\int \mathrm{d}t'\chi_{0\,k,k'}\left(t,t'\right)\left[\Delta_{ext}\left(k',t'\right)\!+\!\Delta_{ind}\left(k',t'\right)\right]
\end{equation}
The BCS mean field theory \cite{bardeen1957theory} corresponds to assuming, 
\begin{equation}\label{ind}
\Delta_{ind}\left(k,t\right)\!=\!\sum_{k'}U_{k,k'}\left\langle c_{-k'\downarrow}\left(t\right)c_{k'\uparrow}\left(t\right)\right\rangle \!=\!\sum_{k'}U_{k,k'}\psi_{k'}\left(t\right)
\end{equation}
where $U_{k,k'}$ is the effective electron-electron interaction potential. Combine Eq.~(\ref{chi} - \ref{ind}), we can express $\chi_{k,k'}\left(t,t'\right)$
of the whole system in the form of operators:
\begin{equation}\label{chichi0}
\boldsymbol{\chi}=\left[1-\boldsymbol{\chi}_{0}\boldsymbol{U}\right]^{-1}\boldsymbol{\chi}_{0}
\end{equation}
where $\boldsymbol{\chi}$ is the operator form of the response function
and $\left[1-\boldsymbol{\chi}_{0}\boldsymbol{U}\right]^{-1}$ is
the inverse of the operator $1-\boldsymbol{\chi}_{0}\boldsymbol{U}$.
Thus we can determine the irradiated superconducting system's response
to the perturbation as long as the noninteracting system's response
function $\boldsymbol{\chi}_{0}$ is obtained. 

We derive the expression of $\boldsymbol{\chi}_{0}\left(t,t'\right)$
with the Keldysh Green function technique \cite{keldysh1965diagram,haug2008quantum}. $\psi_{k}\left(t\right)$
is proportional to the off-diagonal element of the less Green function
defined as $G_{\alpha\beta}^{<}\left(k;t,t'\right)=i\bigl\langle\Psi_{\beta}^{\dagger}\left(t\right)\Psi_{\alpha}\left(t'\right)\bigr\rangle$,
where $\Psi_{1}=c_{k}$ and $\Psi_{2}=c_{-k}^{\dagger}$. With the
perturbation Hamiltonian $V_{ext}$, the Dyson equation of $\boldsymbol{G}^{<}\left(t,t'\right)$
can be obtained with the Langreth Theorem. For noninteracting system,
$\boldsymbol{G}^{<}\left(t,t'\right)$ can be expanded to the first
order of $\Delta_{ext}$ as
\begin{eqnarray}\label{gless}
\bm{G}_{k}^{<}\left(t,t'\right) & = & \bm{G}_{0,k}^{<}+\int \mathrm{d}t_{1}\bm{G}_{0,k}^{r}\left(t,t_{1}\right)\bm{\Sigma}_{k}'\left(t_{1}\right)\bm{G}_{0,k}^{<}\left(t_{1},t'\right)\nonumber \\ [-5pt]
 &  & +\int \mathrm{d}t_{1}\bm{G}_{0,k}^{<}\left(t,t_{1}\right)\bm{\Sigma}_{k}'\left(t_{1}\right)\bm{G}_{0,k}^{a}\left(t_{1},t'\right)
\end{eqnarray}
where
\begin{equation}
\bm{\Sigma}_{k}^{\prime}\left(t\right)=\mbox{\ensuremath{\left(\begin{array}{cc}
0 & \Delta_{ext}\left(k,t\right)\\
\Delta_{ext}^{*}\left(k,t\right) & 0
\end{array}\right)}}
\end{equation}
and $\boldsymbol{G}_{0}^{r}\left(t,t'\right)$ ($\boldsymbol{G}_{0}^{<}$)
is the retarded (lesser) Green function of the noninteracting irradiated
system without the external perturbation. We obtain $\boldsymbol{G}_{0}^{r}$
in the presence of radiation: 
\begin{equation}
G_{0\uparrow\left(\downarrow\right)}^{r}\left(t,t^{\prime}\right)=-i\theta\left(t-t^{\prime}\right)e^{-i\int_{t'}^{t}\mathrm{d}t_{1}\left\{ \pm\tilde{\epsilon}_{k}\pm V_{k}\cos\left(\omega_{0}t_{1}\right)+\frac{i}{2}\Gamma\right\} }
\end{equation}
where $\frac{1}{2}\Gamma$ is the energy level broadening resulting
from the tunneling into the metal substrate. We assume that the self-energy
resulting from the tunneling is nearly a constant around the Fermi
level at energy scale of $\bigtriangleup\varepsilon\sim\Delta$. While
the real part of the self-energy can be absorbed in the electron dispersion,
the imaginary part of the self-energy is approximated to be a constant $-\frac{i}{2}\Gamma$.
$\boldsymbol{G}_{0}^{<}$ can be obtained similarly. By substituting $\boldsymbol{G}_{0}^{r}$ and $\boldsymbol{G}_{0}^{<}$ into Eq.(\ref{gless}), $\boldsymbol{\chi}_{0}\left(t,t'\right)$ is given as
\begin{eqnarray}
\chi_{0\,k,k'}\left(t,t'\right)& = &-i\theta\left(t-t'\right)\int\frac{\mathrm{d}\varepsilon}{2\pi}\Gamma\tanh\left(\frac{\beta\varepsilon}{2}\right)\nonumber\label{chi0_g} \\
& &\cdot e^{-\frac{1}{2}\Gamma\left(t-t'\right)}u_{k}\left(\varepsilon,t\right)u_{k}^{*}\left(\varepsilon,t'\right)\delta_{k,k'}
\end{eqnarray}
with 
\begin{equation}
u_{k}\left(\varepsilon,t\right)=\sum_{n}J_{n}\left(\frac{V_{k}}{\omega_{0}}\right)\frac{e^{i\left(\varepsilon+n\omega_{0}-\tilde{\varepsilon}_{k}\right)t-2i\frac{V_{k}}{\omega_0}\sin\left(\omega_0t\right)}}{\left(\varepsilon+\tilde{\varepsilon}_{k}+n\omega_{0}-\frac{i}{2}\Gamma\right)}
\end{equation}
where $J_{n}\left(x\right)$ is the Bessel function of the first kind and $\delta_{k,k'}$ comes from our assumption that the conservation of momentum is kept during the tunneling process.

With the response function derived, we can determine the conditions of
the superconducting phase transitions. The stability of a linear
system requires all poles of the response function lie on the lower
half plane in the frequency space. The transition condition is determined
when the first pole comes across the real axis during the change of
the system parameters. Driven by periodic radiation field with frequency
$\omega_{0}$, $\chi_{kk'}\left(\omega,\omega'\right)$ and $\Delta_{k}\left(\omega\right)$
can be expressed in matrices defined by $\chi_{kk'}\left(\omega,\omega^{\prime}\right)=\chi_{mn,kk'}\left(\tilde{\omega}\right)\delta\left(\frac{1}{2\pi}\left(\omega-\omega^{\prime}-\left(m-n\right)\omega_{0}\right)\right)$
and $\Delta_{k}\left(\tilde{\omega}+m\omega_{0}\right)=\Delta_{m,k}\left(\tilde{\omega}\right)$,
where $\tilde{\omega}\in\bigl[-\frac{1}{2}\omega_{0},\frac{1}{2}\omega_{0}\bigr]$
and $m,n$ are integers. Equation~(\ref{chichi0}) in the frequency domain shows
that the pole of $\boldsymbol{\chi}\left(\tilde{\omega}\right)$ is
determined by the zero point of $1-\boldsymbol{\chi}_{0}\left(\tilde{\omega}\right)\boldsymbol{U}$.
We can determine eigenvalues and eigenvectors of $\boldsymbol{\chi}_{0}\left(\tilde{\omega}\right)\boldsymbol{U}$.
The stability condition proposed by Bode \cite{bode1945network} requires that for $\tilde{\omega}$
on the real axis, each eigenvalue $x{}_{0i}\left(\tilde{\omega}\right)$
of $\boldsymbol{\chi}_{0}\left(\tilde{\omega}\right)\boldsymbol{U}$
satisfies $\left|x_{0i}\left(\tilde{\omega}\right)\right|\leqslant1$ when $\arg x_{0i}\left(\tilde{\omega}\right)=2\pi$. Thus the transitional point for superconductivity is that for
 the largest $\left|x_{0i}\left(\tilde{\omega}_{c}\right)\right|$
which satisfies $\arg x_{0i}\left(\tilde{\omega}_{c}\right)=2\pi$,
it should be $\left|x_{0i}\left(\tilde{\omega}_{c}\right)\right|=1$. By this
approach we set up a correspondence between the superconducting transitional
temperature $T_{c}$ and the strength of the effective attractive potential $U_{k,k'}$,
as well as the phase frequency $\tilde{\omega}_{c}$ of the gap phase
$\Delta\left(\tilde{\omega}_{c}\right)$. 

In the approach above, we investigate the most unstable mode of the linear system against
perturbation near the phase transition point. To obtain the physical observables at the temperature region far below $T_{c}$, such as the steady-state gap value, nonlinear gap equations should be
applied. Here we make the assumption that the
most unstable solution mode of the linearized equation at the transition
point will correspond to the steady state solution of the nonlinear system. The validity of our assumption
can be assured at the close $T_{c}$ region.

\iffalse $\tilde{\omega}_{c}$ is the oscillation frequency of the gap phase besides the integer or half integer multiple of the radiation frequency $\omega_0$.
Without radiation, it can be proved that $\tilde{\omega}_{c}$  must be zero and Eq.~(\ref{chi0_g}) is consistent with the BCS gap equation. However in the irradiated nonequilibrium state, there is no such conclusion and the occurrence of nonzero $\tilde{\omega}_{c}$ is possible. If the nonzero $\tilde{\omega}_{c}$ state exists, it suggests a new radiation-induced nonequilibrium effect, different from the usual (fractional) ac Josephson effect. In the following paragraphs, we will focus on when the nonzero $\tilde{\omega}_{c}$ state exists with numerical calculations.\fi

The above analysis raises an interesting possibility, i.e., one may find a solution with 
$\tilde{\omega}_{c}\neq0$. There is no a priori reason to believe that  $\tilde{\omega}_{c}$ must be zero or harmonics of the radiation frequency  $\omega_{0}$. Such a solution implies an oscillating state whose frequency is neither zero nor integer (or half integer) multiple of the radiation frequency $\omega_0$. Thus it will be a new radiation-induced nonequilibrium effect, different from the usual (fractional) ac Josephson effect. 
Without radiation, it can be proved that $\tilde{\omega}_{c}$  must be zero and Eq.~(\ref{chi0_g}) is consistent with the BCS gap equation.
However in the irradiated state, there is no such conclusion and the occurrence of nonzero $\tilde{\omega}_{c}$ is possible. In the following paragraphs, we will discuss the existing conditions of this oscillating gap state numerically.

\iffalse We obtain the transitional physical observables with numerical calculations
and find the existence of nonzero phase frequency $\tilde{\omega}_{c}$.
The nonzero $\tilde{\omega}_{c}$ correspond to a dc voltage of $\hbar\tilde{\omega}_{c}/2e$
of the SN junction induced by radiation, while in reverse ac Josephson
effect, the dc voltage of a Josephson junction inspired by radiation
with frequency $f_{0}$ follows the frequency-voltage equation $V_{n}=nhf_{0}/2e$,
where $n$ is integer. The occurrence of this nonzero $\tilde{\omega}_{c}$
suggests a new radiation-induced oscillating state of the gap phase.
In the following paragraphs, we will discuss the generating conditions
and the properties of this oscillating state with numerical calculations.\fi
 
\begin{figure}[tb!]
\includegraphics[width=\columnwidth]{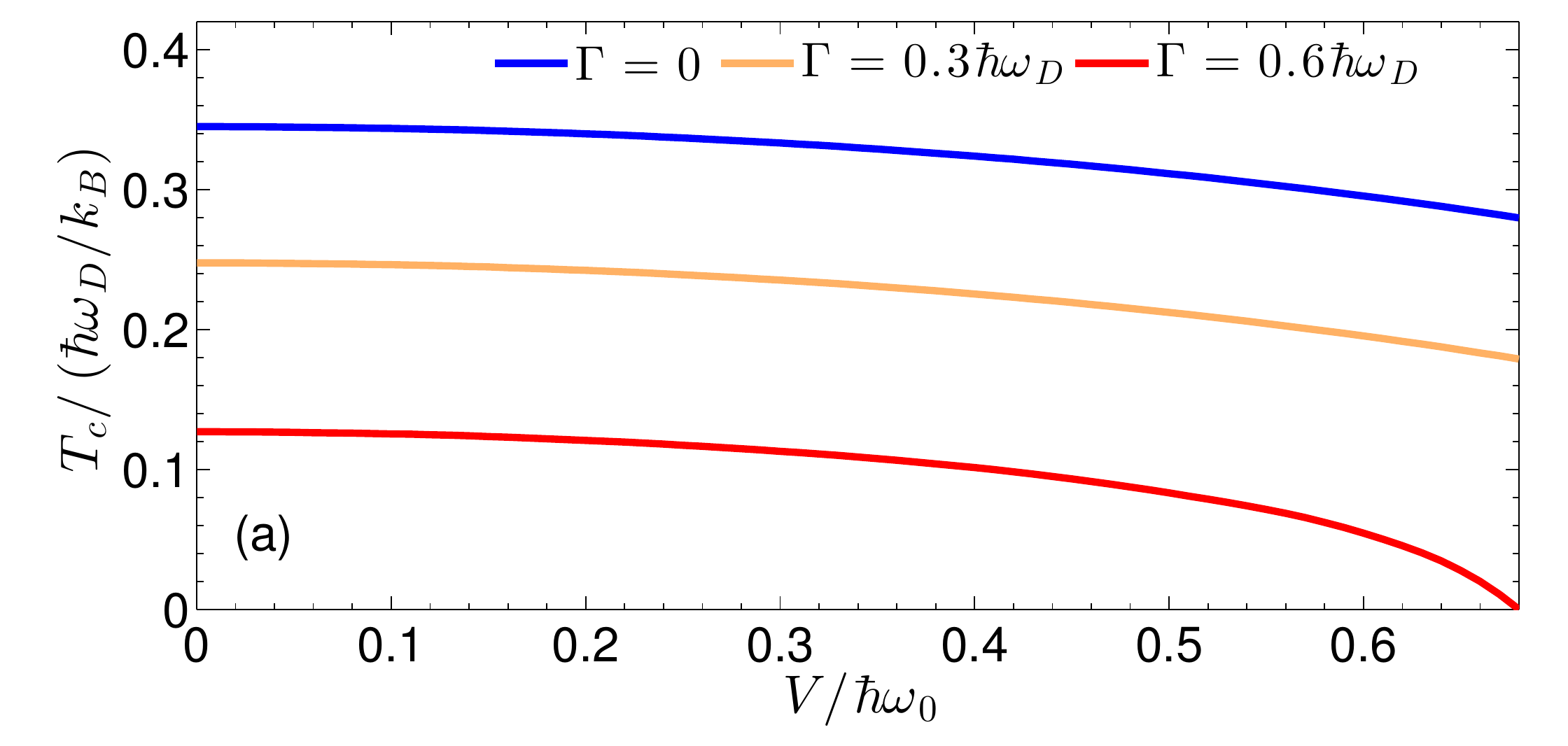} \\% Here is how to import EPS art
\includegraphics[width=\columnwidth]{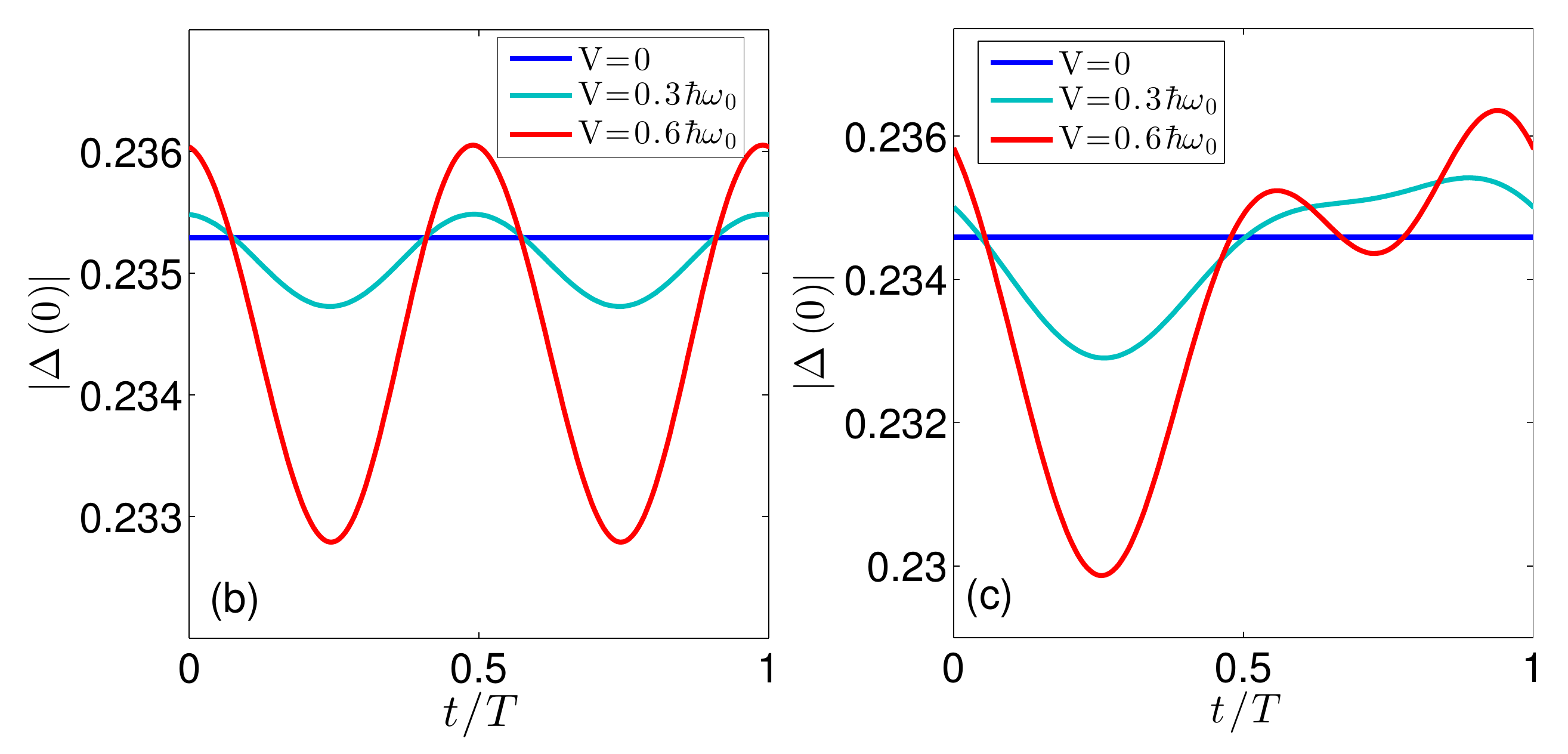}% Here is how to import EPS art
\caption{\label{fig_2} 
(a) The transition temperature $T_{c}$ versus the radiation intensity $V$
for superconductor  with constant DOS at different tunneling amplitudes and $\omega_{0}=0.8\omega_{D}$.
The evolution of the relative gap modulus
 at the Fermi surface with time over one period of the radiation for superconductor (b) at constant DOS or
  (c) at $\rho\left(\tilde{\varepsilon}_{k}\right)=\rho\left(0\right)\left(1+0.08\tilde{\varepsilon}_{k}/\hbar\omega_{D}\right)$ with $\Gamma=0.5$.} 
\end{figure}

First we investigate the effects of radiation on the superconducting
system with constant density of state (DOS) around the Fermi energy. In our calculation, to involve the possible sub-bands motivated by radiations, the effective interaction potential $U_{k,k'}$ is taken to be $U_{k,k'}=-U\theta\left(\hbar\omega_{D}-\left|\tilde{\varepsilon}_{k}-\tilde{\varepsilon}_{k'}\right|\right)$, where $U$ is the magnitude of the attractive potential and $\theta\left(x\right)$ is the step function. We assume
that the radiation frequency is the order of the Debye frequency
$\omega_{D}$, which is in terahertz regime in most BCS superconductors, and that the radiation energy $V_{k}$ and the imaginary
part of self-energy $\Gamma$ due to tunneling are at the order of
magnitude of $\hbar\omega_{D}$. The electron-phonon coupling constant
of the superconducting system assumed as $\lambda=1/U\rho\left(0\right)=0.9$,
where $\rho\left(0\right)$ is the density of state at the Fermi level. 

We show the dependence of the transition temperature $T_{c}$ on the
magnitude of radiation energy $V$ at different tunneling amplitudes
in Fig.~\ref{fig_2}(a). Here
we make the simplification that the radiation energy $V_{k}\approx V$ close to the Fermi energy.
 We can see the transition
temperature $T_{c}$ is suppressed by the radiation. $V\sim T_{c}$
curves at different values of $\Gamma$ show that  tunneling
to the substrate will also suppress the superconductivity.
In this case, the alternating phase frequency $\tilde{\omega}_{c}$ we obtain is
always zero. 

\begin{figure}[tb!]
\includegraphics[width=\columnwidth]{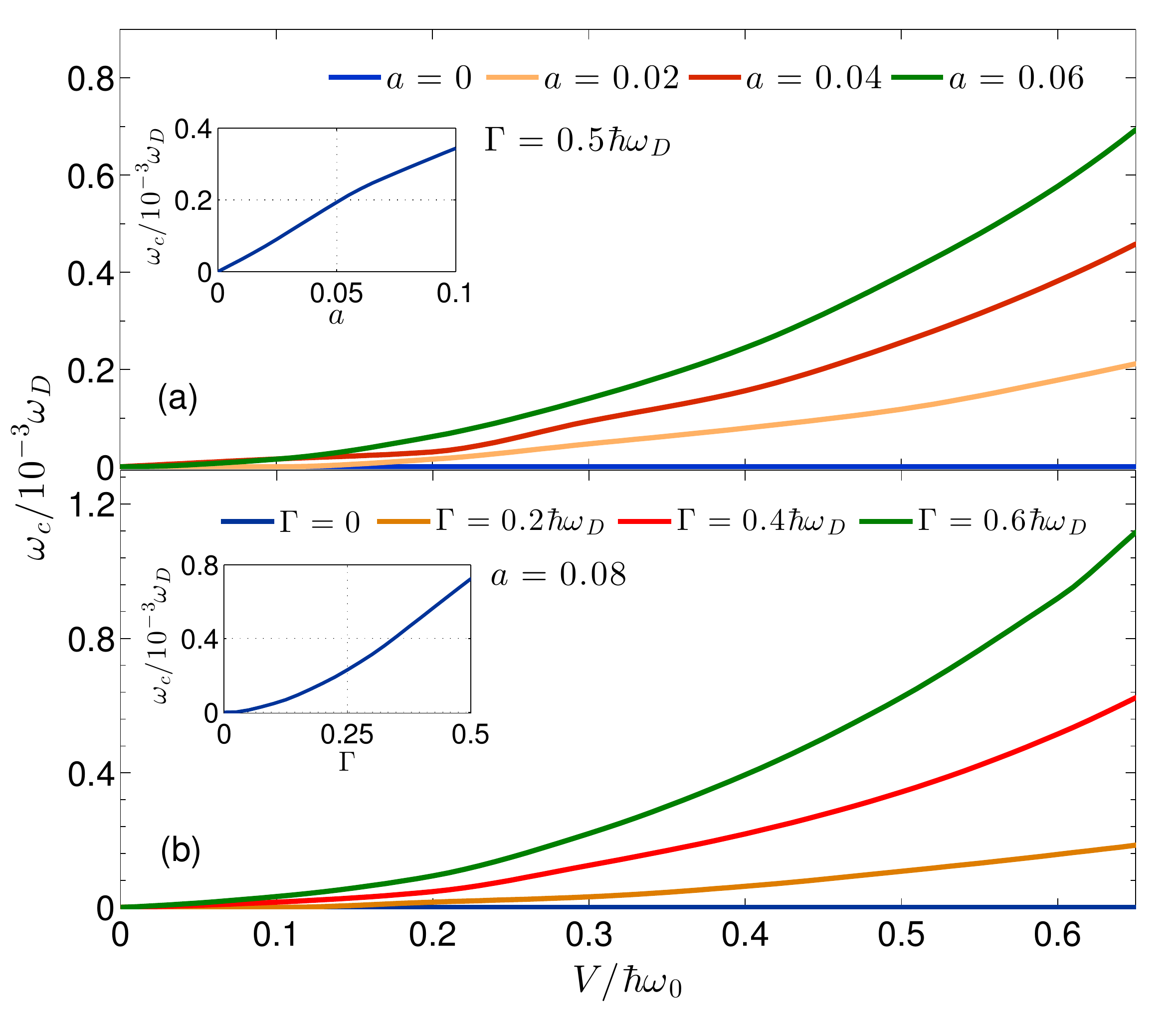}\\% Here is how to import EPS art
\caption{\label{fig_3} 
(a) The alternating phase frequency $\tilde{\omega}_{c}$ versus the radiation intensity $V$ at different values of the slope $a$ of the DOS of $\rho\left(\tilde{\varepsilon}_{k}\right)=\rho\left(0\right)\left(1+a\tilde{\varepsilon}_{k}/\hbar\omega_{D}\right)$ at $\Gamma=0.5\hbar\omega_D$ and $\omega_{0}=0.8\omega_{D}$. Here $\tilde{\omega}_c$ is normalized by $10^{-3}\omega_D$. Insets, dependence of $\tilde{\omega}_{c}$ on $a$ at $V=0.4\hbar\omega_D$. (b)  $\tilde{\omega}_{c}$ versus the radiation intensity $V$ at different values of the tunneling amplitude $\Gamma$ at $a=0.08$ and $\omega_{0}=0.8\omega_{D}$. Insets, dependence of $\tilde{\omega}_{c}$ on $\Gamma$ at $V=0.6\hbar\omega_D$. }
\end{figure}

To search for cases of nonzero $\tilde{\omega}_{c}$,
we investigate the superconducting system with particle-hole
asymmetry near the Fermi surface. We assume the superconductor in our system has a DOS with a linear inclination at the Fermi surface, which is expressed
as $\rho\left(\tilde{\varepsilon}_{k}\right)=\rho\left(0\right)\bigl(1+a\tilde{\varepsilon}_{k}/\hbar\omega_{D}\bigr)$,
where $a$ is the slope and the total range of $\rho$ is limited
to $\left[0.95,1.05\right]$ to avoid unphysical results. In such a system, we observe the occurrence
of the radiation-induced oscillating state. In Fig.~\ref{fig_3}(a), the dependence
of the oscillating frequency $\tilde{\omega}_{c}$ on the radiation intensity $V$at different values of slope is shown. We see that $\tilde{\omega}_{c}$ increase with the DOS slope $a$ as well as the radiation intensity.  $\tilde{\omega}_{c}$
 shows a nearly linear relationship with the asymmetric degree of the
DOS [Fig. \ref{fig_3}(a) inset]. 
$\tilde{\omega}_{c}$ we obtain is in the regime of GHz, three or
more orders of magnitude smaller than the radiation frequency. The
significant difference of magnitude between this alternating 
phase and the radiation frequency could make the experimental observation of the effect easier.

Besides the asymmetry of the DOS, the occurrence of nonzero $\tilde{\omega}_{c}$
also depends on the tunneling amplitudes between the superconductor
and the substrate. Fig.~\ref{fig_3}(b) shows $\tilde{\omega}_{c}$
with respect to the radiation intensity at different values of $\Gamma$. $\tilde{\omega}_{c}$ increase rapidly with the tunneling amplitudes. $\Gamma=0$ corresponds to the bulk superconducting system, where the interactions with the substrate are screened over the length of penetration. At this case, $\tilde{\omega}_{c}$ is always zero, which indicates this radiation induced effect is special to the two-dimensional superconductor.

We calculate the relative modulus of the gap by solving the eigenvectors of the operator $\left(\boldsymbol{U}\boldsymbol{\chi}_{0}\right)^{-1}-1$. The evolution of the gap modulus over one period of radiation at the Fermi surface for systems with constant DOS [Fig.~\ref{fig_2}(b)] and with nonzero $\tilde{\omega}_{c}$
motivated at $a=0.06$ and $\Gamma=0.5$ [Fig.~\ref{fig_2}(c)] are plotted respectively.
 We can see that the radiation induce oscillation of the gap amplitude with the same frequency as the radiation. For system with nonzero $\tilde{\omega}_{c}$, large proportion of the secondary and higher harmonic wave is also excited, as shown in Fig.~\ref{fig_2}(c).

When the system  in Fig.~\ref{fig_1}  with nonzero $\tilde{\omega}_{c}$ is connected to a bulk superconductor or a 2D superconductor with particle-hole symmetry, an unusual alternating Josephson current can be observed experimentally. When irradiated with infrared radiation, the phase difference of the two superconductor is $\phi=\tilde{\omega}_{c}t$. The fast oscillating term due to the infrared radiation is ignored. Thus the tunneling current is $I=I_c\sin\left(\tilde{\omega}_{c}t\right)$, where $I_c$ is affected by the radiation intensity and frequency. Unconventionally, the frequency of this alternating current depends on the radiation intensity and the degree of the electron-hole asymmetry of the 2D superconductor.

In conclusion, we find a radiation-induced oscillating state of
the gap phase in a two-dimensional BCS superconductor with particle-hole asymmetry
coupled to a normal metal substrate. Its oscillating frequency is
determined by the radiation intensity, the tunneling amplitude with
the substrates and the electron-hole asymmetry of the superconducting
system, which is different from the (fractional) ac Josephson effect.
When this system is connected to another superconductor 
in a Josephson junction, alternating current corresponding to this phase occurs. 
\bibliography{sample}% Produces the bibliography via BibTeX.

\end{document}